COMMUNICATION

# Prediction of Flexibility of Metal–Organic Frameworks CAU-13 and NOTT-300 by First Principles Molecular Simulations

Aurélie U. Ortiz,[a] Anne Boutin[b] and François-Xavier Coudert[a]

**Based on first principal calculations, we predict and characterize the flexibility of two aluminium-based Metal–Organic Frameworks (MOFs), CAU-13 and NOTT-300. Both materials have a wine rack topology similar to that of MIL-53(Al), the archetypal breathing MOF, yet their flexibility had not been demonstrated so far.**

Much attention has recently been focused on a fascinating subclass of metal-organic frameworks (MOFs) that behave in a remarkable stimuli-responsive fashion.[1,2] These Soft Porous Crystals (SPCs)[3] demonstrate reversible changes of large amplitude in their crystalline structure and their porous volume under a number of external physical constraints, such as guest adsorption, temperature, or mechanical pressure. Several applications have been proposed to leverage this highly dynamic behaviour for practical applications, including sensing for detecting traces of organic molecules,[4] zero-order kinetics of drugs in long-release single-injection therapies,[5] and specific gas separations.[6,7,8]

One of the most studied modes of flexibility in SPCs is the so-called "breathing" of the MIL-53 family of materials.[9] These materials are formed by unidimensional $M^{III}(OH)$ chains (M = Al, Ga, Cr, In, Sc, Fe, …), linked together by benzenedicarboxylate linkers to form a framework with "wine rack" topology.[10] Owing to this specific topology, in which the inorganic chains act as hinges and the organic linkers at struts,[11] the MIL-53 materials exhibit stimuli-driver crystal-to-crystal transformations between a large pore and a narrow pore structure, with up to 40% change in unit cell volume, upon external stimulation. However, though this eye-catching phenomenon is universal among the MIL-53 family of materials, only two other MOFs with wine rack topologies have been demonstrated to "breathe" so far.[12] The first one is DMOF-1, where the extent of the flexibility is much smaller than MIL-53 (~4% in unit cell volume).[13] The second is MIL-47, an analogue of MIL-53 with VO chains whose flexibility was only recently demonstrated by compression in liquid mercury.[14]

In this Communication, we predict the breathing behaviour of two metal–organic frameworks with wine rack topology, CAU-13 and NOTT-300, whose flexibility has not yet been demonstrated experimentally, by quantum chemical calculations. Furthermore, the methodology exemplified here can be adapted in future theoretical studies of the nature and extent of MOF flexibility.

The first system we investigated is CAU-13, a material similar to MIL-53(Al) in which the benzenedicarboxylate linker is replaced with *trans*-cyclohexanedicarboxylate is a Al(OH)(trans-CDC).[15] The as-synthesized CAU-13 structure features a wine rack topology, with two water molecules inside the pores which can be removed upon activation, and exhibits microporosity upon $N_2$ adsorption. However, the flexibility of CAU-13 has not yet been established, though some structure transition is expected due to its close resemblance to the breathing MIL-53 framework. We thus characterized the mechanical behaviour of CAU-13 by two series of quantum chemical calculations. First, we characterized the elastic properties of the structure of the water-free CAU-13 framework, using *ab initio* quantum mechanical calculations in the density functional theory approach with localized basis sets (CRYSTAL09 code[16]), the B3LYP hybrid exchange-correlation functional,[17] and dispersion corrections.[18] This methodology has been well validated on both rigid[19] and flexible MOFs.[12] The full stiffness matrix of CAU-13, containing 21 second-order elastic constants due to its triclinic nature, are reported in Table S2. Tensorial analysis[20] allowed us to extract from those the directional Young's modulus, shear modulus, linear compressibility and Poisson's ratio, whose extremal values are summarized in Table I. It appears clearly that CAU-13 bears all the signatures of SPCs:[12] very large anisotropy in its elastic properties (a factor of 5.6 in Young's modulus, and ~25 in shear modulus), and the existence of deformation modes of low modulus, in particular a minimal shear modulus of ~2 GPa. Moreover, CAU-13 shares one more trait of wine rack-based structures, namely a large negative linear compressibility. All these characteristics indicate that it should be able to exhibit "breathing" behaviour.

|  | $E_{min}$ | $E_{max}$ | $A_E$ | $G_{min}$ | $G_{max}$ | $A_G$ | $\beta_x$ (TPa$^{-1}$) | $\beta_y$ (TPa$^{-1}$) | $\beta_z$ (TPa$^{-1}$) | $v_{min}$ | $v_{max}$ |
|---|---|---|---|---|---|---|---|---|---|---|---|
| **CAU-13** | 25.23 | 140.12 | 5.6 | 1.98 | 50.55 | 25.5 | –2.0 | 162.1 | –31.8 | –1.48 | 3.07 |
| **NOTT-300** | 3.19 | 120.06 | 37.7 | 0.8 | 32.55 | 40.09 | 9.3 | 9.3 | 3.9 | –1.99 | 2.72 |

**Table I.** Minimal and maximal values, as well as anisotropy, of Young's modulus $E$, shear modulus $G$, linear compressibility $\beta$, and Poisson's ratio $v$ for the CAU-13 and NOTT-300. Anisotropy of $X$ is denoted by $A_X = X_{max} / X_{min}$.

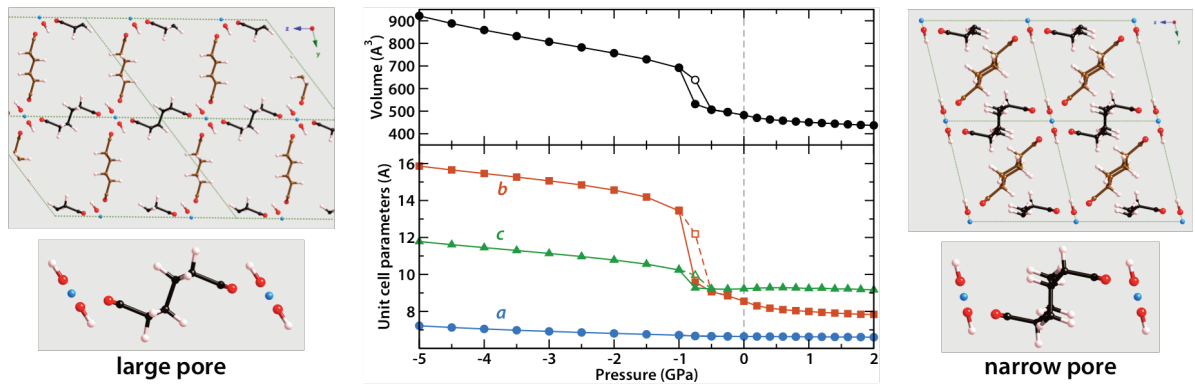

**Figure 1.** Centre: evolution of CAU-13 unit cell volume and cell parameters $a$, $b$ and $c$ as a function of mechanical pressure (stress). Right: structure of CAU-13 at 0 GPa ("narrow pore" form). Left: structure of CAU-13 under a pressure of –1 GPa ("large pore" form).

This prediction, based on the elastic properties only, needs to be confirmed with calculations of the actual deformation of the material. Though structural transitions can be triggered by either gas adsorption or mechanical stress, they are two facets of the same phenomenon.[21] Because direct simulation of adsorption in the osmotic ensemble is too costly for the current *ab initio* methods, we propose here an alternative by using *in silico* tension and compression pressure experiments, representative of mechanical or adsorptive stress.

The pressure dependence of CAU-13's unit cell parameters is presented in Fig. 1. A structural transition can clearly be seen in the range –0.5 to –1 GPa: it corresponds to an increase of 43% in unit cell volume, mainly due to expansion along the $b$ axis. This transition corresponds in particular to a straightening of the $a,a$-CDC linker (insets in Fig. 1; Fig. S1) between two Al(OH) chains. The "large pore" structure obtained upon tension has higher symmetry, with $I2/m$ space group (compared to $P$-1 for the original structure), and larger pore volume. There is thus a complete parallel with the know case of MIL-53(Al) at low temperature (below 200 K), where the narrow pore form is the thermodynamically stable guest-free form.[22] The critical stress at which the transition occurs (–0.5 GPa) is in the same range as typical outward adsorption stress in microporous materials near saturation uptake (which can reach into the GPa range).[23,24] This suggests the predicted large pore structure of CAU-13 could be observed experimentally by adsorption at relatively high gas pressure (near , or adsorption of bulky guest molecules at modest pressure. Indeed, the breathing of CAU-13 has very recently been observed under adsorption of xylenes, though the structure of {xylene,CAU-13} phases has not been solved.[25]

In a second time, we studied the mechanical behaviour of NOTT-300,[26] a recently synthesized MOF based on 1D Al(OH) chains, linked together by tetracarboxylate linkers to form square-shaped channels (Fig. 2 and Fig. S1). Though similar in topology to the MIL-53 family, the connection between Al atoms in the framework of NOTT-300 are through *cis*-$\mu_2$-OH groups. Yang et al., who did not observe any sign of flexibility of the material upon solvent removal, guest adsorption, or variation of the temperature, attributed this apparent rigidity to NOTT-300's *cis* configuration of the $\mu_2$-OH groups.[26] Yet, the framework of NOTT-300 shows a wine rack topology, which usually yields high flexibility.[11,12]

The elastic behaviour of NOTT-300, reported in Table I (full elastic tensor in Table S3) shows the same telltale signs of structural flexibility as CAU-13 and MIL-53 materials, and in particular a very low shearing modulus in the [1 –1 0] direction, $G_{min}$ = 0.8 GPa. We confirmed this behaviour by studying the structural response of the material under tetragonal shear, as depicted in Fig. 2. The deformation of the material is continuous and reversible, in the orthorhombic $I2_12_12_1$ space group (the relaxed NOTT-300 is quadratic, with space group $I4_122$). No first-order structural transition happens under shearing. Below 1 GPa, we observe a gradual compression of the framework, occurring with breathing-like motion: $a$ decreases while $b$ increases. At 1 GPa, the NOTT-300 framework has reached a "narrow pore" structure (Fig. 2), with reduced pore volume and surface area (Table S2). At that point, the material undergoes hardening, with much lower compressibility at higher pressure. This shows that, contrary to the expectations, there is a large degree of structural flexibility in NOTT-300, with a possible reduction of unit cell volume of ~24%, despite the difference in the connectivity of the framework (*cis*-$\mu_2$-OH vs. *trans*-$\mu_2$-OH).

Finally, while we demonstrated above that the mechanically-induced breathing of NOTT-300 could be achieved by imposing a shearing stress, we wanted to check whether it could also be triggered by isotropic compression, such as hydrostatic fluid compression in a diamond anvil cell. This could not be solved by zero-Kelvin energy minimization, in which an isotropic compression cannot break the quadratic symmetry of the NOTT-300 framework. We thus performed first principles molecular dynamics (FPMD; full details in ESI) of the material at 300 K, to see if the breathing could occur under isotropic pressure in the presence of thermal fluctuations. FPMD results confirm the occurence of a spontaneous rupture of the quadratic symmetry of NOTT-300 at pressures of 700 MPa and higher, with contraction into the narrow pore state (Fig. 3). This transition is reversible, with the larger pore state being recovered if pressure is released. Moreover, at intermediate pressure (500 MPa), we do not observe the transition in our limited simulation time, although some softening of the framework is evident from the larger fluctuations of unit cell volume: this is similar to the pressure-induced softening observed in ZIF-8 as being a precursor to the structural transition.[27]

We thus predict the possibility of mechanically-induced breathing of NOTT-300, which has not been observed experimentally so far. The mechanical compression required for NOTT-300 breathing, between 500 and 750 MPa, is higher than the pressures observed for MIL-53 and MIL-47 (55 and 137 MPa, respectively[14,21]) under mechanical compression. Whether this "narrow pore" state can be triggered by guest adsorption is another question. In order to shed some light into this, we loaded the narrow pore structure with 11 water molecules (full loading), and relaxed the atomic positions and unit cell parameters. Instead of reverting back to the large-pore structure, as in the guest-free phase, the NOTT-300 material converged to a third phase structure (in ESI), with an opening angle of $\theta = 74°$ ($a = 12.9$ Å and $c = 16.6$ Å), intermediate between narrow-pore ($\theta = 54°$) and large-pore ($\theta = 90°$). This intermediate structure is reminiscent of that observed in breathing MIL-53(Fe), and suggests that breathing in NOTT-300 might possibly be triggered by adsorption, in addition to mechanical stress. Future work is needed to assess the conditions of guest pressure at which this phenomenon would happen.

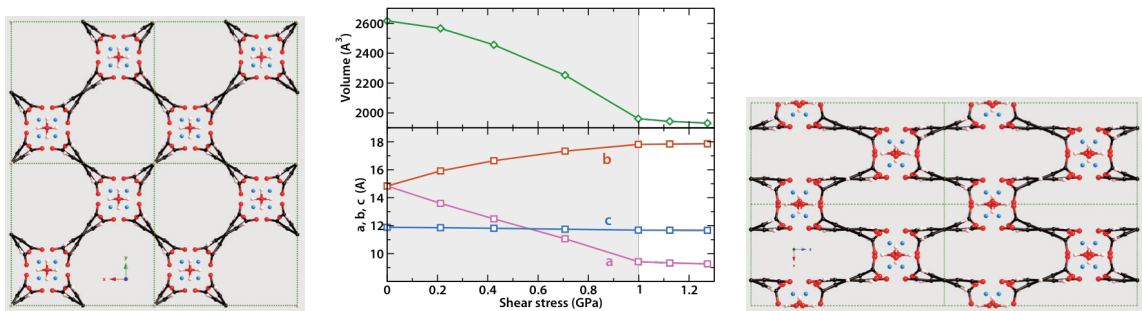

**Figure 2.** Left: structure of NOTT-300 at P = 0 GPa ("large pore" form). Centre: evolution of NOTT-300 unit cell volume and parameters as a function of tetragonal shear stress. Right: structure of the "narrow pore" form of NOTT-300, under shear stress of 1 GPa.

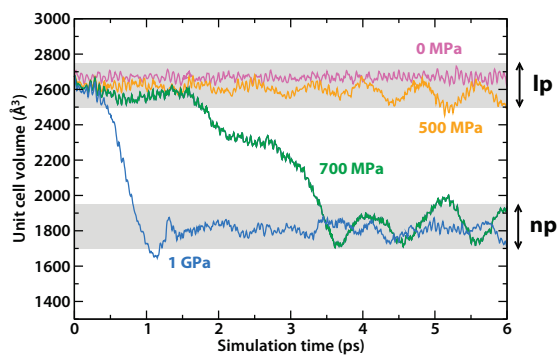

**Figure 3.** Time evolution of the unit cell volume of NOTT-300, during FPMD at 300 K and various values of pressure

Finally, these observations pave the way for a comprehensive computational study of NOTT-300's behaviour as a function of temperature and pressure, in particular free energy methods to investigate the thermodynamics of the breathing in NOTT-300 at non-zero temperature, to check whether the gradual nature of the phenomenon, observed in quantum chemical calculations, is retained at all temperatures or switches at some temperature to a behaviour closer to bistable MIL-53 and CAU-13.

To conclude, we presented a complete methodology for the computational characterization and prediction of flexibility in MOFs, combining three different types of first principles molecular simulations to give a broad view of the flexibility of SPCs. In the future, this methodology may be used to screen newly synthesized MOFs for flexibility, since framework flexibility does not always manifest itself immediately by *in situ* variable-temperature XRD or adsorption of simple gases.

We acknowledge funding from Agence Nationale de la Recherche (project ANR-2010-BLAN-0822) and computing time on HPC platforms by GENCI (grants i2014086114 and i2014087069).

## Notes and references

[a] Institut de Recherche de Chimie Paris, CNRS – Chimie ParisTech, 11 rue Pierre et Marie Curie, 75005 Paris, France.

[b] Département de Chimie, École normale supérieure, CNRS-ENS-UPMC, 24 rue Lhomond, 75005 Paris, France.

† Electronic Supplementary Information (ESI) available: full details of the calculations, stiffness matrices, structures calculated in CIF format.